# Fermilab Proton Accelerator Complex Status and Improvement Plans

*Vladimir Shiltsev*

*Fermi National Accelerator Laboratory, PO Box 500, Batavia, IL 60510, USA*

*shiltsev@fnal.gov*

*Abstract*

Fermilab carries out an extensive program of accelerator-based high energy particle physics research at the Intensity Frontier that relies on the operation of 8 GeV and 120 GeV proton beamlines for a n umber of fixed target experiments. Routine operation with a world-record 700kW of average 120 GeV beam power on the neutrino target was achieved in 2017 as the result of the Proton Improvement Plan (PIP) upgrade. There are plans to further increase the power to 900 – 1000 kW. The next major upgrade of the FNAL accelerator complex, called PIP-II, is under development. It aims at 1.2MW beam power on target at the start of the LBNF/DUNE experiment in the middle of the next decade and assumes replacement of the existing 40-years old 400 MeV normal-conducting Linac with a modern 800 MeV superconducting RF linear accelerator. There are several concepts to further double the beam power to >2.4MW after replacement of the existing 8 GeV Booster synchrotron. In this article we discuss current performance of the Fermilab proton accelerator complex, the upgrade plans for the next two decades and the accelerator R&D program to address cost and performance risks for these upgrades.

Keywords: High intensity accelerators; Neutrino physics; Fermilab; Space-charge effects.

PACS numbers: 29.20.dk, 29.25.-t, 29.27.bd

# 1. Introduction: Overview of the Fermilab accelerator complex

Fermilab accelerator complex (see Fig. 1) is one of the largest in the world and consists of 16 km of accelerators and beamlines, two high power targets, several low power target stations, many experiment and service buildings, etc. It delivers beams of protons and secondary particles. All the protons come from a 750 keV H- RFQ, followed by a 400 MeV normal-conducting pulsed linac injecting to 8 GeV rapid-cycling-synchrotron (RCS) Booster, which is largely an original construction ca 1960's. The Booster combined function dipole magnets operate in a 15 Hz resonant circuit, which sets a fundamental clock for the complex. However, historically not all cycles could be loaded with protons due to imitations from injection, extraction, and RF system and beam loss. The next machine downstream in the complex is the Recycler, a 3.3 km 8 GeV storage ring made from permanent magnets. Originally built for storage and accumulation of low intensity antiproton beams during the Tevatron Collider Run II (2001-2011) [1], the Recycler is now used to stack high intensity protons for loading into the 120 GeV Main Injector synchrotron which is located in the same tunnel. Circumference of the Recycler is sufficient to accommodate six batches of 84 Booster bunches each; however, through the technique called "slip-stacking", six more batches can be injected in the machine in addition to six slightly decelerated original batches, making the total of 12 batches. The kinds of batches travel at slightly different velocities and "slip" with respect to each other until they overlap and at that moment they are transferred to the Main Injector and accelerated to maximum energy (more details on the "slip-stacking" method can be found in [1]). The Fermilab proton accelerator complex supports a number of experiments – e.g., the 400 MeV Linac beam is sent to the Mucool Test Area, 8 GeV protons from the Booster are supplied to the 8 GeV Booster Neutrino Beam (BNB), ANNIE, MicroBooNE, MiniBooNE, MITPC, SciBath, ICARUS (near future), and SBND (future), and to muon experiments "muon g-2" and "Mu2e" (near future). The 120 GeV proton beam from the Main Injector supports neutrino experiments at NuMI (MINOS+, MINERvA, NOvA) and LBNF/DUNE in the future, as well other fixed target experiments SeaQuest, LArIAT and at the Test Beam Facility. See Ref.[2] for detailed information on these experiments.

In May of 2014, the Particle Physics Project Prioritization Panel (P5), advisory to the Office of High Energy Physics in the US Department of Energy, released a report [3], which identifies the top priority of the domestic intensity frontier high-energy physics for the next 20-30 years to be a high energy neutrino program to determine the mass hierarchy and measure CP violation, based on the Fermilab accelerator complex which needs to be upgraded for increased proton intensity. The current long baseline neutrino program utilizes the Neutrinos from the Main Injector (NuMI) beam line that was built to provide protons for the MINOS experiment, located in the Soudan Mine in Minnesota, 725 km away. Later, the NOvA experiment was built 810 km away in Ash River, MN. It also uses the NuMI beam line, but it is built 14.6 mrad off axis, producing a narrower neutrino energy spread, resulting in an improved resolution for the CP violating phase and mass hierarchy. The physics goal set forth by the P5 report is: "…a mean sensitivity to CP violation of better than 3σ… over more than 75% of the range of possible values of the unknown CP-violating phase $δ_{CP}$". To this end, a new beam line and experiment are being planned. The beam line is the Long

Baseline Neutrino Facility (LBNF) [4] at FNAL and the new experiment is the Deep Underground Neutrino Experiment (DUNE) [5], located in the Sanford Underground Research Facility (SURF) near Lead, South Dakota, 1300 km away from Fermilab. This will be a truly international collaboration, including contributions from 150 institutions in 27 countries. The P5 physics goals require about 900 kt·MW·years of exposure (product of the neutrino detector mass, average proton beam power on the neutrino target and data taking period). Assuming a 40 kton Liquid Argon detector, this would take over 50 years at the 400 kW beam intensity which was typical when the program was first conceived. For this reason, a series of accelerator upgrades toward the eventual goal of multi MW proton beam power have been undertaken and planned.

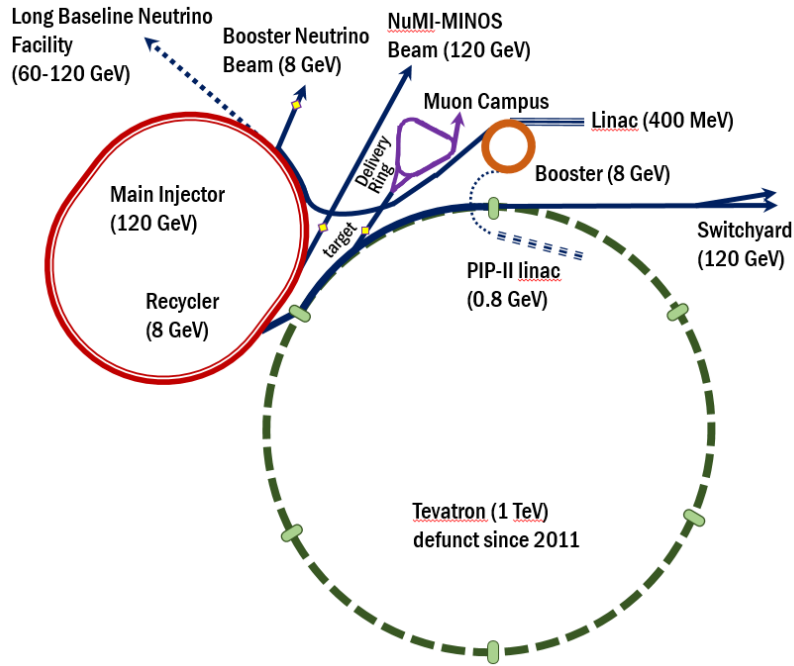

FIG.1: Fermilab accelerator complex includes 400 MeV H- pulsed normal-conducting RF linac, 8 GeV proton Booster synchrotron, 8 GeV Recycler storage ring that shares tunnel with 120 GeV proton Main Injector synchrotron, and a 3.1 GeV muon Delivery Ring. A number of beamlines connect the accelerators, bring the beams to fixed targets and to various high energy physics experiments. Most notable future additions (dashed lines) include the LBNF beam line for DUNE and 0.8 GeV CW-capable SRF PIP-II linac located inside the Tevatron ring and corresponding beamline for injection into the Booster (see in the text).

## 2. Current upgrade activities - Proton Improvement Plan (PIP) and PIP-I+

The goal of the Proton Improvement Plan (PIP) campaign is to maximize the proton output from the existing complex with an ultimate goal of routine operation at the 700 kW of 120 GeV beam power from the Main Injector. The key elements of PIP are reduction of losses and upgrades of the pulsed RF hardware in the Booster to allow beam to be accelerated on all 15 Hz cycles. This goal has recently been achieved and the total proton output from the Booster achieved 2e17 protons per hour. In addition, commissioning of the 6+6 batch slip-stacking in the Recycler allowed to reduce the Main Injector cycle time to 1.33s from 2.2s during the MINOS/Collider Run II era. Due to these improvements, in early 2017 the Main Injector achieved a world-record of 716 kW average proton beam power over one hour to the NuMI beam line – see Fig.2. On the way, the operations team increased the number of batches slip-stacked in the Recycler in steps (just 6 batches in late 2014, then 2+6, 4+6 and, finally, 6+6 batches in mid-2016). At each step, the increase in intensity was followed by tuning for efficiency and minimization of losses. Installation of the Recycler beam collimation system and commissioning of a more efficient beam feedback system to control coherent instabilities during the slip-stacking process were the keys to the record.

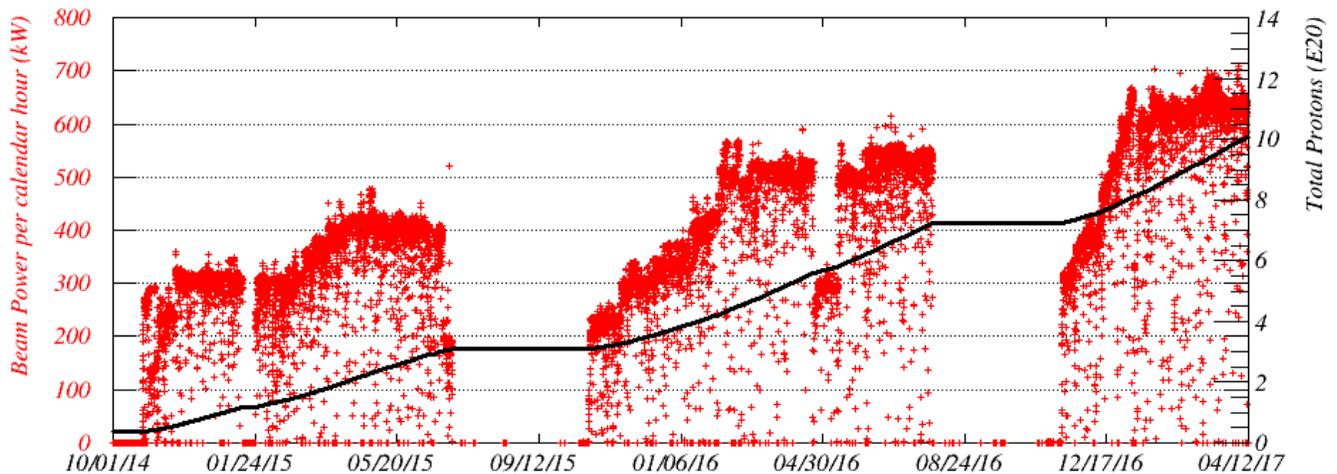

FIG.2: Three years record 2014-2017 of the hourly average 120 GeV proton beam from the Main Injector synchrotron (courtesy P.Adamson).

Fermilab accelerator operations team is currently considering a PIP-I+ campaign aimed at increasing the NuMI beam power to 0.9-1 MW level without major changes to the existing accelerator complex. The campaign needs to be finished prior to the construction of the PIP-II linac (see next chapter) and includes upgrades of the NuMI target station to be robust up to 1 MW, series of improvements to reduce losses in the Booster to be able to operate with some 30% higher number of protons per pulse (PPP) – see Fig.3a, increasing the Booster rate from 15 Hz to 20Hz and corresponding modification of the Recycler slip-stacking RF system and the Main Injector ramp to allow faster 1.1s cycle.

## 3. Proton Improvement Plan – II (PIP-II)

In the current configuration, it's unlikely that significantly more beam could be injected into the Booster. The preceding machine, the Linac, can provide more than $6 \cdot 10^{12}$ 400 MeV protons per pulse but the efficiency of the Booster drops with the intensity increase and, hence, the operations team needs to keep the PPP at the present level of about $4.3 \cdot 10^{12}$. The beam losses occur at the injection energy (some 3-4%) and at the "transition energy" of $\gamma=5.48$ (another 0.3-0.5%). At higher intensities both the beam emittance and beam losses grow beyond acceptable levels due to strong space-charge forces – see Fig.3b. The space-charge effects scale inversely with the squared beam size $\sigma^2$ and are proportional to beam current (number of particles in the bunch $N$) and the time the beam spends at low energies ($\gamma$). From the latter argument, fast acceleration with a gradient of about 5-20 MeV/m in RF Linacs is advantageous compared to that in the rapid cycling synchrotron (RCS) rings which is usually about 0.002-0.01 MeV/m. Correspondingly, the maximum RCS bunch intensity $N$ increases as the $\beta\gamma^2$ of the beam injected from the linear accelerator (see e.g., [6]).

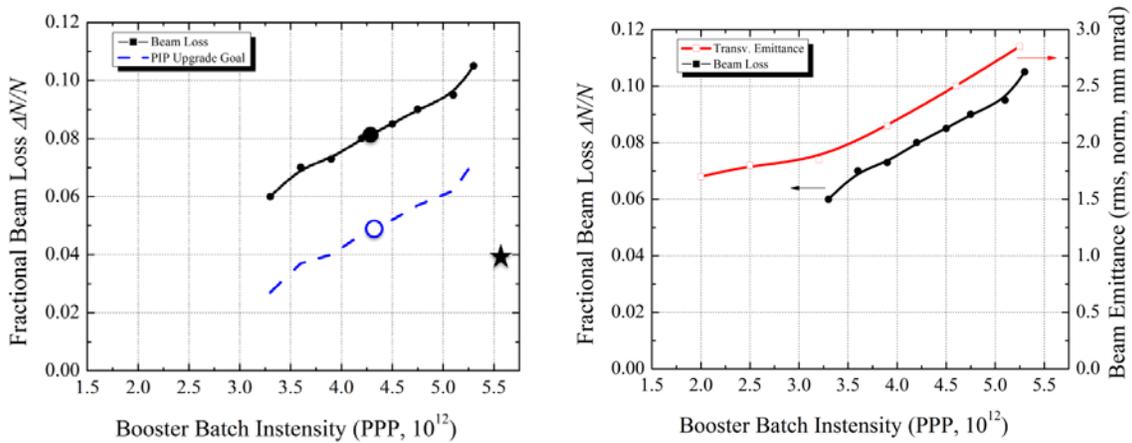

FIG.3: Space-charge effects in the Fermilab's 8 GeV proton Booster synchrotron vs the total intensity of the batch of 84 bunches in the accelerator: a) right – fractional intensity losses at present (solid black line), as expected after the Proton Improvement Plan (dashed blue line) and the operational goal of the PIP-I+ upgrade (star); b) left - fractional beam loss integrated over the entire Booster acceleration cycle at present (black, left axis) and beam emittance (red line, right axis, adapted from [22]).

The key feature of the proposed Proton Improvement Plan-II (PIP-II) project [7] is, therefore, to replace the lab's existing 40-year-old 400 MeV normal-conducting RF pulsed linear accelerator with a new one800 MeV based on superconducting RF cavities, capable of CW operation. It will allow an increase of the 120 GeV proton beam power available to the new LBNF beamline to 1.2 MW. In addition, the Booster rate will be increased from 15 to 20 Hz, allowing full MI beam power to be achieved at the lower energy of 60 GeV (vs current 120 GeV), and 80 kW of beam for the 8 GeV neutrino program in addition to ~100 kW of proton beam power available at 800 MeV with arbitrary bunch structure - see full list of the PIP-II key technical parameters in the Table 1.

Table 1. Beam parameters of PIP and PIP-II upgrades.

| Performance Parameter | PIP | PIP-II | |
|---|---|---|---|
| Linac Beam Energy | 400 | 800 | MeV |
| Linac Beam Current | 25 | 2 | mA |
| Linac Beam Pulse Length | 0.03 | 0.6 | msec |
| Linac Pulse Repetition Rate | 15 | 20 | Hz |
| Linac Beam Power to Booster | 4 | 18 | kW |
| Booster Protons per Pulse | $4.3\times10^{12}$ | $6.5\times10^{12}$ | |
| Booster Pulse Repetition Rate | 15 | 20 | Hz |
| Booster Beam Power @ 8 GeV | 80 | 160 | kW |
| Beam Power to 8 GeV Program (max; MI @ 120 MeV) | 32 | 80 | kW |
| Main Injector Protons per Pulse | $4.9\times10^{13}$ | $7.6\times10^{13}$ | |
| Main Injector Cycle Time @ 60-120 GeV | 1.33 | 0.7-1.2 | sec |
| LBNF/NuMI Beam Power @ 60-120 GeV | 0.7 | 1.0-1.2 | MW |
| LBNF Upgrade Potential @ 60-120 GeV | NA | >2 | MW |

The PIP-II project got so called "CD-0 approval" from the US DOE in 2015 and is scheduled for completion in 2025-2026. Currently, the project team carries out an extensive R&D program focused on reduction of the technical risk and the total project cost via development of the superconducting (SC) RF cavities, research towards significant improvement of the quality factor of the SRF cavities and construction and test of the PIP-II low-energy part ("front-end"). The PIP-II SC RF cavity development is multi-faceted as cavities of five different types, operating at three different frequencies are needed: the 162.5 MHz Half-Wave Resonators (HWRs) – their design is complete and the first cryomodule is in production; the 325 MHz Single-Spoke Resonators (SSR1) – their design is mostly complete and production started; and the design of the 650 MHz High-Beta resonators is also well advanced. All these developments involve close collaboration with Indian institutions [8].

The high-$Q_0$ SRF studies have recently discovered that Nitrogen doping during the Nb cavity surface processing [9] more than doubles the cavity's quality factor $Q_0$ and, thus, reduces the required cryogenic capacity. It was also found that fast cooling of the cavities (which operate at 2K) enhance the magnetic flux expulsion out of the SC cavity [10] and improves $Q_0$, too. The front-end Linac test facility [11] will demonstrate in practice the two most challenging PIP-II design elements: very low energy transition from "room-temperature" RF acceleration to "cold" SC RF at the of 2.1 MeV and a 162.5 MHz CW beam chopper. The PIP-II Injector Test facility is being built in collaboration with ANL, LBNL and SNS and two Indian institutions (BARC, IUAC) and has recently achieved beam acceleration through its 2.1 MeV CW RFQ.

### 4. R&D towards future multi-Megawatt upgrade of the FNAL accelerator complex

Fermilab has recently started the conceptual development of the next upgrade of its accelerator complex to multi-Megawatt beam power levels after 2030 and began the corresponding R&D. The key objectives of the upgrade, tentatively called PIP-III [12], are: i) attainment of more than 2.4 MW beam power for the LBNF/DUNE neutrino program (that is twice the PIP-II goal); ii) replacement of the Booster as it is currently a bottleneck toward higher intensities [13]; iii) affordable cost of the upgrade; iv) usage of the PIP-II proton linac and the Main Injector synchrotron; v) development of multi-MW beam targets [14].

The very first of these conditions calls for either double the PPP out of the Booster replacement machine in the case that the Recycler ring is kept in operation with slip-stacking, or even quadruple it in the expected case of inability to operate the Recycler with the slip-stacking due to the process's intrinsically lossy bunch manipulations. In general, charged particle accelerators, particularly, the high intensity proton ones, are very complex and several, rather than single, improvements are needed to get to provide significant increase in their performance [15]. In the PIP-III case, the beam losses are the main obstacle to enable the multi-MW power upgrade. E.g., assuming usual tolerable radiation levels ~ 1W/m in the accelerator enclosures, the total losses must be kept under 500 W in the new Booster which will be operating at the 320 kW extracted beam power level (i.e., the fractional beam loss should not exceed either <0.15% at the extraction energy or ~2% at injection); and be less than 3kW in the Main Injector with 2.4 MW beams (that corresponds to fractional losses <0.12% at the top energy or ~2% at the injection energy). These numbers are extremely low and very challenging at the anticipated record high beam intensities [16] – compare with the present level of the beam losses of about 3-5% in the Booster and in the Main Injector under the conditions of 2-4 times lower PPP.

We plan to explore several approaches for such an upgrade: i) construction of a SRF linac which would extend the PIP-II energy to 8 GeV; ii) 8 GeV RCS which can handle ×4 stronger space-charge forces [17]; iii) a hybrid option of, e.g., 2 GeV Linac and 8 GeV RCS. The advantage of the linac option is that it would provide copious power at 8 GeV, both for ancillary program and so the high energy program could be run at full power at lower Main Injector energies; the major challenge will be exploration of new techniques for significant reduction of the construction cost per GeV compared to that of a number of the SRF-based accelerator facilities [18]. Technical challenges of the H- ion injection into the Recycler or Main Injector at 8 GeV also needs careful evaluation.

The advantage of the RCS is that the requisite minimal beam-loss performance has been demonstrated in the J-PARC 3 GeV RCS [19]. Disadvantageous would be limited proton beam power at 8 GeV. Even if a new RCS is built, the normalized emittance will continue to be limited by the acceptance of the Main Injector, so we would not benefit from a large physical aperture like that of the J-PARC RCS. This means that additional measures against the space-charge effects should be undertaken. One option would be to increase the energy of the linac and, therefore the RCS injection energy beyond the 800 MeV energy of the PIP-II linac, if that can be done in a cost efficient manner. Another option would be to mitigate the effect of the space charge using non-linear integrable optics [20] or electron lenses [21].

It has been shown in [20] that contrary to conventional linear focusing accelerator optics lattices, single particle dynamics exhibits a significantly higher degree of stability in certain integrable nonlinear lattices that would permit mitigation of many of the space charge restrictions that limit beam intensity. Practical realization of such lattices is possible with, e.g., special types of nonlinear magnets (see, e.g. [22]). Numerical studies of space charge effects in such nonlinear lattices with intense bunches indicate strong suppression of beam halo formation [23]. Another opportunity to create such nonlinear element is to use high intensity magnetized electron beam of an electron lens [21].

Electron lenses with Gaussian transverse beam current distribution similar to those developed for the head-on beam-beam compensation in the Tevatron and RHIC [21] can be used for direct space-charge compensation (SCC) [24]. The method of SCC [25] calls for placement of stable negative charge of electrons on the orbit of a high intensity proton beam, such that the transverse density profiles of both charges match and compensate each other as illustrated in figure 12. Besides electron lenses, such a configuration can be obtained via accumulation of ionization electrons in strong magnetic traps, "electron columns" [26].

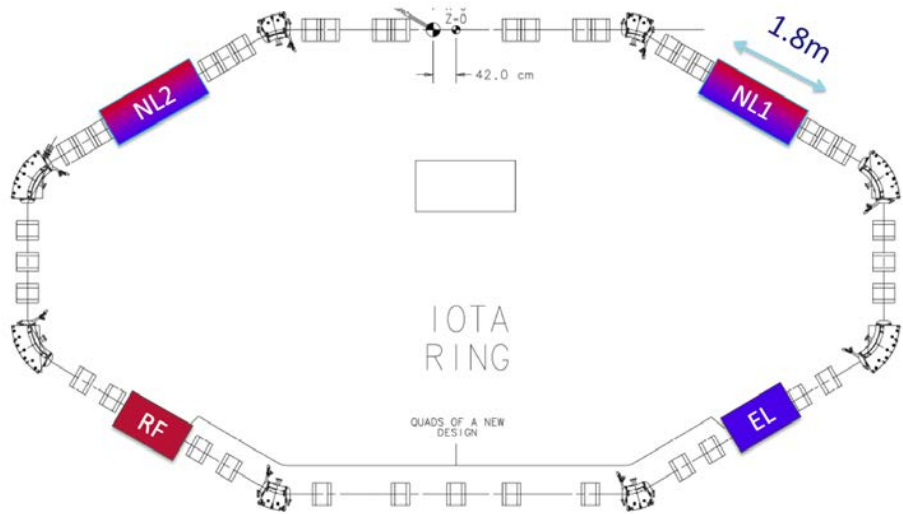

FIG.4: IOTA ring layout with two non-linear magnets and an electron lens for the space-charge compensation (adapted from [22]).

All these novel ideas will be experimentally tested at the Integrable Optics Test Accelerator (IOTA) ring – see Fig.4 – at the Fermilab Accelerator Science and Technology (FAST) facility [22] which, besides the ring, consists of its 150 MeV electron injector and 70 MeV/*c* (2.5 MeV kinetic energy) proton RFQ injector. The R&D program at the 40-m circumference IOTA ring requires development and installation of the corresponding non-linear magnets and electron lens, and operation with either 150 MeV electrons for an initial test of the integrable optics or 70 MeV/c high brightness proton beam for the space-charge mitigation with IO and/or with electron lens and with an electron column. The IOTA/FAST facility is under construction now, the 52 MeV electron beam out of the SRF photoinjector has been commissioned in 2016, the first circulating electrons in the IOTA are expected in 2018 and the first protons in 2019-2020. The experimental research program at IOTA is outlined in Ref.[22], it will require a very well controlled and stable accelerators [27], and advanced beam diagnostics comparable or surpassing in its variety the suit of instruments developed for the Tevatron collider program [28]. It is expected that the experimental research at IOTA will also be of relevance for future colliders [29, 30]. The first tests and studies with currently available beams at the IOTA/FAST facility has already began [31].

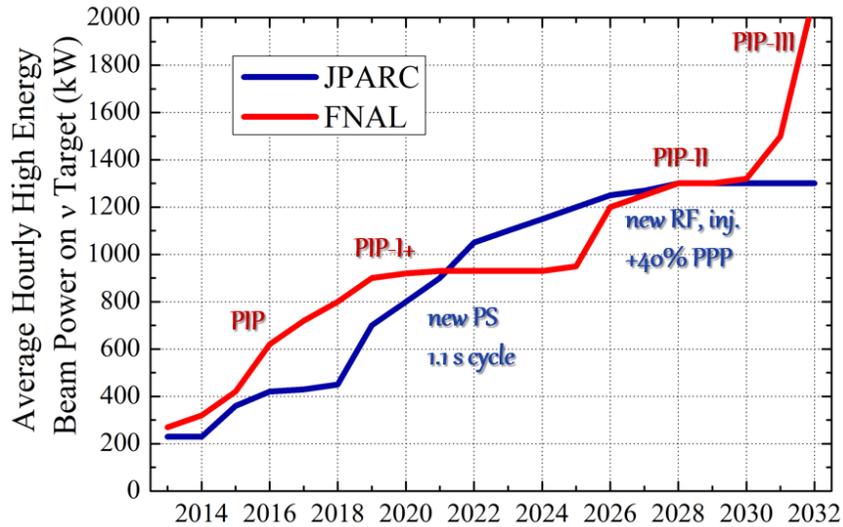

FIG.5: Performance to date and planned hourly average high energy proton beam power on neutrino targets out of the Fermilab and JPARC accelerators (2013-2032, see text).

**5. Summary and discussion**

Fermilab aspires to leadership in the Intensity Frontier of high energy particle physics. The accelerator-based neutrino research program relies on the operation of the 120 GeV proton beamlines for NuMI at present and for the LBNF/DUNE experiment in the middle of the next decade. Routine operation with a world-record 700kW of average high-energy beam power on the neutrino target has been achieved in 2017 and a further increase to 900-1000 kW level is anticipated after a series of relatively minor improvements to the existing accelerator complex in the coming years. The next major upgrade of the FNAL accelerator complex, the PIP-II project, aims at 1.2MW beam power on the neutrino target at the start of the LBNF/DUNE experiment. It assumes replacement of the existing 400 MeV pulsed normal-conducting Linac with a modern 800 MeV superconducting RF linear accelerators. We also explore several concepts to replace existing 8 GeV Booster and further double the beam power to >2.4MW. All these upgrade plans will allow to keep up world record in the high-energy proton beam power for neutrino research and stay ahead of the major competitor, Japan's JPARC [19] – see Fig.5. An extensive accelerator R&D program with significant international contribution has been launched to address cost and performance risks for these upgrades: the PIP-II Injector Test facility, development of cost-effective SRF cavities and experimental R&D program at the IOTA ring to demonstrate novel space-charge mitigation methods.

**Acknowledgments**

Author would like to acknowledge fruitful discussions with and contributions from Phil Adamson, David Bruhwiler, Mary Convery, Paul Derwent, Stephen Holmes, Ioanis Kourbanis, Valery Lebedev, Sergei Nagaitsev, William Pellico, Eric Prebys, Alexander Romanenko, Cheng-Yan Tan, Alexander Valishev, and Robert Zwaska. This paper mostly follows the talks on the status, challenges and plans for the Fermilab accelerators given at the "*Neutrino-2016*" conference (July 4-9, 2016, London, UK) [32] and at the *2017 April Meeting* of the American Physical Society (January 28-31, 2017, Washington, DC, USA) and I would like to thank the organizers of these events for inviting me to give these presentations. Fermilab is operated by Fermi Research Alliance, LLC under Contract No. De-AC02-07CH11359 with the United States Department of Energy.